\documentclass[11pt,twoside]{article}

\usepackage{asp2006}
\usepackage{graphicx}

\begin{document}
\title{Properties and Evolution of AGN Jet Ridge Lines \\ The Caltech-Jodrell Bank Flat-Spectrum Sample}
\author{Marios Karouzos\altaffilmark{1}, Silke Britzen\altaffilmark{1}, Andreas Eckart\altaffilmark{2,1}, and Anton J. Zensus\altaffilmark{1,2}}
\altaffiltext{1}{Max-Planck-Institut f\"ur Radioastronomie, Auf dem H\"ugel, 53121, Bonn, Germany}

\altaffiltext{2}{I.Physikalisches Institut, Universit\"at zu K\"oln, Z\"ulpicher Str. 77, 50937, K\"oln, Germany}

\begin{abstract} We investigate the jet morphology and kinematics of a statistically complete radio-loud AGN sample in terms of the gamma-ray properties of the sources. Gamma-ray detected AGN dominate the high end of the jet apparent speed distribution of the total sample. Gamma-variable sources show stronger evolution in their jet morphology. A 5.1\% of the sources show large ($>15$ degrees) swings in their jet ejection angle.
\end{abstract}


\section{Introduction}
We study the jet kinematics and jet ridge line evolution of the Caltech-Jodrell Bank flat-spectrum (CJF) sample. The sample consists of 293 radio-loud active galaxies \citep{Taylor1996}, spanning a large redshift range ($z_{avg}=1.254$; $z_{max}=3.889$). Each source has at least 3 epochs of radio interferometric (VLBA) observations and has been imaged and studied kinematically \citep{Britzen2007b,Britzen2008}.

\section{Radio Jets in the CJF}
\subsection{Apparent Speeds and $\mathbf{\gamma}$-ray Properties}
\begin{figure*}[!ht]
\begin{center}
  \includegraphics[width=0.7\textwidth]{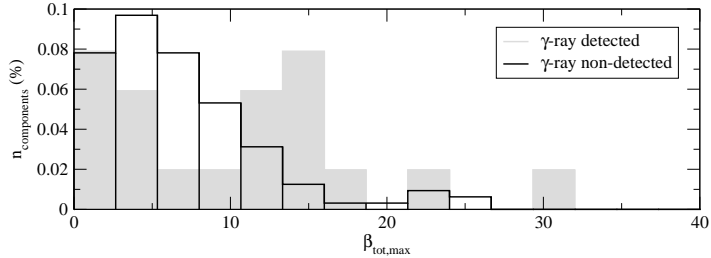}
   \caption{Distribution of the maximum total apparent speed $\beta_{tot,max}$ for $\gamma$-ray detected (shaded) and non-detected (non-shaded) sources (data from \citealt{Britzen2008}, \citealt{Hartman1999}, and \citealt{Abdo2009}).}
  \label{fig:gamma_speeds}
\end{center}
\end{figure*}

 We correlate the kinematic information \citep{Britzen2008} with the $\gamma$-ray properties of the sample. In total 25 sources have been detected in $\gamma$-rays. 18 CJF sources are included in the 3 month bright AGN source list of the Fermi-LAT \citep{Abdo2009}, with 7 additional sources detected by EGRET \citep{Hartman1999}. The maximum of the apparent speed distribution for the $\gamma$-ray non-detected sources is $3\le\beta_{tot,max}\le5$. For the detected sources the distribution shows two maxima ($\beta_{tot,max;1}=0$ and $\beta_{tot,max;2}=15$). The detected sources dominate the high speed tail of the total distribution (Fig. \ref{fig:gamma_speeds}).
\subsection{AGN Jet Ridge Lines}
We define the ridge line of a jet as the line that linearly connects all component positions at a given epoch. We build the jet ridge lines for all CJF sources (for S5 1803+784, see Britzen et al. 2009, A\&A, in press) and study their properties and evolution across all available epochs.

\subsubsection{Ejection Angle Evolution}
We define the angle between the innermost jet component and the core, $theta$, as the approximate ejection angle of the jet (the core being at the beginning of the coordinates system). We study the change of this angle across all available epochs. Most sources exhibit a maximum ejection angle change $<5^{\circ}$. 5.1\% of the sources show ejection angle changes $>15^{\circ}$, with two sources at $\Delta\theta_{max}\approx45^{\circ}$.
\subsubsection{Jet Ridge Lines and $\mathbf{\gamma}$-ray properties}
We quantify the jet ridge line evolution as the total component position change across all epochs. $\gamma$-ray variable sources show stronger ridge line evolution than non-variable ones. Sources with stronger ridge line evolution tend to be more luminous in the $\gamma$-rays.

\section{Conclusions}
We find that $\gamma$-ray detected sources show higher apparent speeds than non-detected ones. $\gamma$-variable sources show stronger jet ridge line evolution. $5.1\%$ show considerable jet ejection angle changes ($>15^{\circ}$). Such jet ejection angle changes might be connected to precessing or helical jets.
\acknowledgements
M. Karouzos was supported for this research through a stipend from the International Max Planck Research School (IMPRS) for Astronomy and Astrophysics. M.K. wants to thank C.S. Chang for proofreading this poster and for insightful comments on various points of this work.

\begin{thebibliography}
\bibitem[Abdo et al. (2009)]{Abdo2009}Abdo, A. A., Ackermann, M., Ajello, M., et al. 2009, \apj, 700, 597A
\bibitem[Britzen et al. (2008)]{Britzen2008}Britzen, S., Brinkmann, W., Campbell, R. M., Gliozii, M., Readhead, A. C. S., Browne, I. W. A., \& Wilkinson, P. 2007, A\&A, 476, 759]{Britzen2007b}
\bibitem[Britzen et al. (2007b)]{Britzen2007b}Britzen, S., Vermeulen, R. C., Campbell, R. M., et al. 2008, A\&A, 484, 119
\bibitem[Hartman et al. (1999)]{Hartman1999}Hartman, R.C., Bertsch, D. L., Bloom, S. D., et al. 1999, ApJS, 123, 79
\bibitem[Taylor et al. (1996)]{Taylor1996}Taylor, G. B., Vermeulen, R.C., Readhead, A. C. S., Pearson, T. J., Henstock, D. R., \& Wilkinson, P. N. 1996, ApJS, 107, 37
\end{thebibliography}

\end{document}